\newcommand\beq{\begin{eqnarray}}
\newcommand\eeq{\end{eqnarray}}
\renewcommand\d{\partial}

\renewcommand{\theequation}{\thesection.\arabic{equation}}

\newcommand\la{\langle}
\newcommand\ra{\rangle}
\newcommand{\cl}{\centerline}

\newcommand{\qvec}{\mbox{\boldmath $q$}}
\newcommand{\pvec}{\mbox{\boldmath $p$}}
\newcommand{\kvec}{\mbox{\boldmath $k$}}
\def\slash#1{\rlap/{#1}}
\def\kslash{\slash{\mkern-1mu k}}    
\def\pslash{\slash{\mkern-1mu p}}
\def\qslash{\slash{\mkern-1mu q}}
\def\uslash{\slash{\mkern-1mu u}}    
\def\dmu{\partial^{\mu}}

\documentstyle[12pt]{article}

\begin{document}
\setlength{\baselineskip}{0.2in}

\begin{titlepage}
\noindent
\hfill MSUCL-870

\hfill January, 1993

\noindent

\setlength{\oddsidemargin}{0.6in}
\setlength{\evensidemargin}{0.6in}
\setlength{\textwidth}{6.5in}
\setlength{\textheight}{9in}
\setlength{\parskip}{0.05in}
\vspace{36pt}

\vspace{5pt}
\cl{\Large{\bf Octet Baryons at Finite Temperature:}}
\cl{\Large{\bf QCD Sum Rules vs. Chiral Symmetry}}
\par
\par\bigskip
\par\bigskip
\vspace{0.7cm}

\cl{Yuji KOIKE}
\cl{\em National Superconducting Cyclotron Laboratory,
Michigan State University}
\cl{\em East Lansing, MI 48824-1321, USA}
\vspace{1in}

\cl{\bf Abstract}
Correlators of the octet baryons in the hot pion gas are studied
in the framework of the QCD sum rule.
The condensates appearing in the OPE side of the correlators
become $T$-dependent through the interaction with thermal pions.
We present an explicit demonstration
that the $O(T^2)$-dependence of the condensates
is completely
compensated by the change of the pole residue and
the $\pi + B \rightarrow B'$ scattering effect in the spectral functions.
Therefore the baryon masses are constant to this order,
although $\la \bar{u} u\ra_T \simeq \la \bar{u}u\ra_0(1-T^2/8f_\pi^2)$,
which is consistent with the
chiral symmetry constraint by Leutwyler and
Smilga.

\end{titlepage}

\newpage
\section{Introduction}
\setcounter{equation}{0}
\renewcommand{\theequation}{\arabic{section}.\arabic{equation}}

Modification of the hadronic properties
at finite temperature has been receiving greater attention
in connection with the relativistic heavy ion collisions such as RHIC and LHC.
Although the finite-$T$
properties of various mesons have been extensively studied
in the literature using QCD Sum Rules (QSR)\,\cite{BS,HKL,HKL2}
 and effective theories for QCD\,\cite{Pis,HK,Shu,DEI},
baryon properties have not been seriously investigated except for
the nucleon\,\cite{LS,El,AZ}.
In principle, the change of baryon masses should manifest itself in
the yield modification and the threshold behaviour of
lepton pairs coming from the
baryon-antibaryon annihilation in the relativistic
heavy ion collisions.  Therefore it is a pressing  issue to have a QCD
prediction for the finite-$T$ behavior of the baryon correlators.

Since the work by Shifman, Vainstein and Zakharov\,\cite{SVZ}, the
QCD sum rule
method
has been extensively used as a systematic tool to study various
resonance properties based on QCD (see \cite{RRY} for a review).  Especially,
the masses of octet and decuplet baryons have been well reproduced
in terms of the vacuum condensates\,\cite{If,RRY2,BI,EPT}.
Hence it is worth while to extend
this strongly QCD motivated phenomenology to study finite-$T$ behavior
of baryon correlators.
  In this paper we shall
analyze the correlators of
the octet baryons (N, $\Lambda$, $\Sigma$ and $\Xi$) at
finite temperature.  The source currents
we will investigate
are the following
interpolating fields which have been used in QSR:
\beq
\eta^N(x) &=& \epsilon^{abc}u^a(x)C\gamma_\mu u^b(x)
\gamma_5\gamma_\mu d^c(x),\label{eq1}\\[5pt]
\eta^{\Lambda}(x) &=&
\sqrt{2\over3} \epsilon^{abc}\left[u^a(x)C\gamma_\mu s^b(x)
\gamma_5\gamma_\mu d^c(x) -
d^a(x)C\gamma_\mu s^b(x) \gamma_5\gamma_\mu u^c(x) \right],\label{eq2}\\[5pt]
\eta^{\Sigma}(x)
&=& \epsilon^{abc}u^a(x)C\gamma_\mu u^b(x) \gamma_5\gamma_\mu s^c(x),
\label{eq3}\\[5pt]
\eta^{\Xi}(x)
&=& -\epsilon^{abc}s^a(x)C\gamma_\mu s^b(x) \gamma_5\gamma_\mu u^c(x),
\label{eq4}
\eeq
where $a$, $b$, $c$ are the color indices and $C$ is the charge conjugation
matrix.  Although one can choose other sets of source currents for the octet
baryons, it has been known that the above combination gives
the best description
of the octet baryon masses.

A useful quantity for studying the finite-$T$ behavior of baryons
is the retarded
correlation function for the above fields\,\cite{AGD}:
\beq
\Pi^R_{\alpha\beta}(\omega,\qvec,T)=i\int\,d^4x\,e^{iqx}
\theta(x_0)\la
\eta_\alpha(x) \bar{\eta}_\beta(0) + \bar{\eta}_\beta(0)\eta_\alpha(x) \ra_T,
\label{eq5}
\eeq
where $q=(\omega, \qvec)$ is the external four momentum,
$\la{\cal O}\ra = Tr\left({\cal O}e^{-H/T}\right)/Z$ ($Z=Tr(e^{-H/T})$)
 is the thermal average for the operator ${\cal O}$
with $H$ being the QCD hamiltonian, and $\alpha$ and $\beta$ are the spinor
indices of the interpolating fields (\ref{eq1})--(\ref{eq4}).
The retarded correlator satisfies the dispersion relation
with an appropriate spectral function $\rho(\omega,\qvec,T)$:
\beq
\Pi^R_{\alpha\beta}(\omega,\qvec,T)=\int_{-\infty}^{\infty}\,du\,
{ \rho_{\alpha\beta}(u,\qvec,T) \over {u-\omega-i\epsilon} }.
\label{eq6}
\eeq
In the QCD sum rule method, one applies the operator product expansion (OPE)
to the left hand side of the correlator in the deep Euclidean region
$Q^2 = -q^2 \rightarrow \infty$ and adjusts the
 resonance parameters (resonance masses, pole residues,
width of the resonances, continuum threshold, etc) in the spectral function
$\rho$ so as to reproduce the OPE side of the correlator.
This procedure provides us with an expression for the resonance parameters in
terms of the condensates.
Therefore, in the QSR approach, the change of the condensates at finite-$T$
naturally causes a response of the baryons as a change of their properties
at finite-$T$.
For example, one might naively expect
that the nucleon mass will drop at finite temperature
as the chiral order parameter decreases\,\cite{AZ}.

On the other hand, Leutwyler and Smilga\,\cite{LS}
considered the nucleon correlator
in a thermal pion gas and showed that the nucleon mass does not have
$O(T^2)$-dependence:  In the thermal pion gas,
the $T$-dependence of the nucleon correlator is associated with
the pion-nucleon forward scattering amplitude.  It becomes zero
in the soft pion limit (Adler's zero) and thus the real part of the self-energy
becomes zero at the pole position of $T=0$.
The same statement is also true for all the octet baryons.

The purpose of this paper is to demonstrate how this seeming
contradiction can be reconciled in the framework of the finite-$T$
QCD sum rules.
We organize the QSR at $T\neq 0$ for the nucleon in the
dilute pion gas\,\cite{HKL}.  This approximation should be valid
at relatively low temperature ($T\le 150\ {\rm MeV}$) well inside the confined
phase.  In this framework, the condensates appearing in the OPE side of
the correlators receive $T$-dependence through the pion matrix elements
of the same operators.
We pay particular attention to the effect of the $\pi + N \rightarrow N$
scattering in the phenomenological side of the QSR as was suggested
by Eletsky\,\cite{El}.
Then it can be shown that the $O(T^2)$-dependence of the condensates
is exactly compensated by this scattering
and the mass does not have $O(T^2)$-dependence as was found by Leutwyler
and Smilga.  Thus the naive expectation motivated by the Ioffe's mass
formula for the nucleon, $M_N(T)\simeq
\left(-2(2\pi)^2\la\bar{u}u\ra_T\right)^{1/3}$,
does not work at finite-$T$.  The situation is analogous
for the other octet baryons.

The construction of this paper is the following: In section 2, after
briefly summarizing our finite-$T$ QSR for the case of the nucleon
along the line of \cite{HKL}, we present an explicit demonstration
that the nucleon mass does not have a $O(T^2)$-dependence in the QSR.
The discussion for the hyperons is similar.  So we will
not repeat it, but present some of the formulas in the Appendix.
In section 3, we shall analyze the baryon correlators in the pion gas
in terms of PCAC without using QCD sum rules.  We shall see that the
octet baryon correlators at finite-$T$ can be written
in terms of the correlators at $T=0$ with $T$-dependent coefficients
and our OPE expression derived in section 2 is consistent with those
relations.
This explains why the pole positions of the octet baryon correlators
do not have $O(T^2)$-dependence.
Section 4 is devoted to a brief summary and outlook.

\section{QCD Sum Rules for Nucleon at $T\neq 0$}
\setcounter{equation}{0}
\renewcommand{\theequation}{\arabic{section}.\arabic{equation}}
\subsection{OPE at $T\neq 0$}

We shall first discuss
the QCD side of the correlator.   For simplicity, we study
the nucleon at rest, i.e. $\qvec={\bf 0}$.  In the region,
$Q^2 = -q^2 =-\omega^2 >0$, the retarded correlator (\ref{eq5}) is
identical to the causal correlator:
\beq
\Pi_{\alpha\beta}(\omega,\qvec,T)=i\int\,d^4x\,e^{iqx}
\la T \left(
\eta_\alpha(x) \bar{\eta}_\beta(0) \right) \ra_T.
\label{eq201}
\eeq
At $T\neq 0$, $\Pi_{\alpha\beta}(q,T)$ can be decomposed into the three
scalar components:
\beq
\Pi_{\alpha\beta}(q,T) = \Pi_1(q,T)\delta_{\alpha\beta} +
\Pi_2(q,T)\qslash_{\alpha\beta} +
\Pi_3(q,T)\uslash_{\alpha\beta},
\label{eq202}
\eeq
where $u_\mu$ is the average four-flow velocity of the medium equal to
$u_\mu=(1,0,0,0)$ in the rest frame.  In the deep Euclidean region $Q^2
\rightarrow \infty$, one can apply OPE to $\Pi(q,T)$, which, for
$\Pi_2$ as an example, can be
schematically written as,
\beq
\Pi_2(q,T) &=& \sum_i C_i^{\mu_1\cdot\cdot\cdot\mu_s}(q,\mu)
{\cal O}^{i(d,s)}_{\mu_1\cdot\cdot\cdot\mu_s}(\mu)\nonumber\\[5pt]
&=& {-1 \over{64\pi^4}} q^4{\rm ln}(Q^2) +
\left( \la {\cal O}^{(4,0)}\ra_T + {q^{\mu}q^{\nu} \over q^2 }
\la{\cal O}^{(4,2)}_{\mu\nu}\ra_T \right){\rm ln}(Q^2) \nonumber\\[5pt]
&+&
{ 1 \over q^2 } \left( \la{\cal O}^{(6,0)}\ra_T
 + {q^{\mu}q^{\nu} \over q^2 }
\la {\cal O}^{(6,2)}_{\mu\nu} \ra_T+
{ q^{\mu}q^{\nu}q^{\lambda}q^{\sigma} \over q^4 }
\la{\cal O}^{(6,4)}_{\mu\nu\lambda\sigma}\ra_T \right) + \cdot\cdot\cdot,
\label{eq203}
\eeq
where the $i$-th local
operator ${\cal O}^{i(d,s)}_{\mu_1\cdot\cdot\cdot\mu_s}(\mu)$ renormalized
at the scale $\mu$
has dimension-$d$ and spin-$s$ with $s$ Lorentz indices and
$C_i^{\mu_1\cdot\cdot\cdot\mu_s}(q,\mu)$ is the corresponding Wilson
coefficient.
As a complete set of the local operators in (\ref{eq203}), one can always
choose symmetric and traceless operators with respect to all Lorentz
indices.  We will hereafter assume this symmetry condition for
all nonscalar operators.
In the above equation the following two features peculiar to
$T\neq 0$ are implemented\,\cite{HKL}:

(i) $T$-dependence of the correlators appears only as a thermal average of
the local operators in the OPE as a consequence of the QCD factorization.
This is indeed natural if we note that
such a
soft effect should be ascribed to the condensates $\la {\cal O}^i \ra_T$
as long as the temperature is low enough
compared to the separation scale $\mu$, i.e. $T \ll \mu \ll Q$.

(ii) At $T\neq 0$, there is no Lorentz invariance due to the presence of
the thermal factor $e^{-H/T}$, and hence nonscalar operators
survive as condensates:
\beq
\la {\cal O}^i_{\mu_1\cdot\cdot\cdot\mu_s}(\mu)\ra_T =
(u_{\mu_1}\cdot\cdot\cdot u_{\mu_s} - {\rm traces})a^i(\mu,T).
\label{eq204}
\eeq
At relatively low temperature in the confined phase, the
system can be regarded as a noninteracting gas of Goldstone bosons (pions).
In this approximation, $T$-dependence of the condensates
can be written as
\beq
\la {\cal O}^i \ra_T \simeq \la {\cal O}^i \ra
+ \sum_{a=1}^3 \int { d^3p \over {2\varepsilon(2\pi)^3 }}
\la \pi^a(\pvec)|{\cal O}^i|\pi^a(\pvec)\ra n_B(\varepsilon/T),
\label{eq205}
\eeq
where $\varepsilon = \sqrt{\pvec^2 + m_\pi^2 }$, $a$ denotes the isospin
index, $n_B(x)=[e^x -1]^{-1}$ is the Bose-Einstein distribution function,
and $\la\cdot\ra$ is the usual vacuum average.
Here we have used the covariant normalization for the pion state:
$\la \pi^a(\pvec)|\pi^b(\pvec')\ra =2\varepsilon(2\pi)^3 \delta^{ab}
\delta^3(\pvec-\pvec')$.
Thus we need pion matrix elements
of the local operators appearing in the
OPE to carry out the finite-$T$ sum rule.
For the scalar operators ${\cal O}^{(d,0)}$, we can apply
the soft pion theorem
\beq
\la \pi^a(\pvec)|{\cal O}|\pi^b(\pvec)\ra
={-1\over f_\pi^2}\la0|\left[{\cal F}_5^a, \left[{\cal F}_5^b, {\cal O}
\right]\right]|0\ra + O\left({m_\pi^2 \over {\Lambda_{HAD}^2}} \right),
\label{eq206}
\eeq
where $\Lambda_{HAD}$ is a typical hadronic scale of $O(1\,{\rm GeV})$ and
${\cal F}^a_5$ is the isovector axial charge defined by
\beq
{\cal F}_5^a = \int\,d^3x\, \bar{q}(x)\gamma_0\gamma_5 {\tau^a \over 2}
q(x).
\label{eq207}
\eeq
Pion matrix elements of the nonscalar operators are associated with
the pion structure functions measurable in the hard processes (deep
inelastic scattering, Drell-Yan, direct photon production etc.).
Experimentally, however, only twist-2 part of these
matrix elements are known to some extent, and therefore we have to await
future precise measurements of the twist-4 pion structure functions
to carry out the satisfactory QSR at finite-$T$.
Pion matrix elements of these operators read
\beq
\la \pi(p)|{\cal O}^i_{\mu_1\cdot\cdot\cdot\mu_s}|\pi(p)\ra \sim
(p_{\mu_1}\cdot\cdot\cdot p_{\mu_s} - {\rm traces} )A_i(\mu),
\label{eq208}
\eeq
and therefore the contribution of nonscalar condensates becomes
the effect of $O(T^4)$ or higher in $T$ ($m_\pi \sim T$ is assumed) as
is easily seen by inserting (\ref{eq208}) into (\ref{eq205}).   In \cite{HKL}
we found that these effects can be neglected below $T =160$ MeV.
We shall consistently ignore these condensates in this work.
For the same reason, $\Pi_3(q,T)$ also becomes higher order with respect
to temperature.
The effect of the heavier resonances (K, $\eta$ etc) was also found
to be negligible at $T\le 160$ MeV because of the
suppression coming from the distribution function $\sim
e^{-m_K/T}$\,\cite{HKL}.

Applying the soft pion theorem to scalar operators appearing in the
OPE of the nucleon correlators, we get
\beq
\Pi^N_1(q,T)&=& {1\over 4\pi^2}\la\bar{u}u\ra
\left(1-{\zeta\over 8}\right)
q^2{\rm ln}(Q^2),\label{eq209}
\\[10pt]
\Pi^N_2(q,T)&=& {-1\over 64\pi^4}q^4{\rm ln}(Q^2)
- {1 \over 32\pi^2}
\la{\alpha_s \over \pi} G^2\ra{\rm ln}(Q^2) -
{2\la\bar{u}u\ra^2 \over 3 q^2},\label{eq210}
\eeq
where $\zeta=(T^2/f_\pi^2)B_1(m_\pi/T)$ with
\beq
B_1(z)= {6 \over \pi^2} \int_z^\infty \,dy\, \sqrt{y^2-z^2}{1\over{e^y -1}}.
\label{eq216}
\eeq
In (\ref{eq209}) and (\ref{eq210}) we note the following points:

(i) Although we discarded the terms proportional to $m_u$ and $m_d$ because
of their smallness, we included the pion mass correction coming from the
Bose-Einstein distribution, since it appears in the
form of ${m_\pi/T}$ in $B_1$.  $B_1$ approaches 1 at $m_\pi \ll T$,
while it is strongly suppressed as $B_1 \sim e^{-m_\pi/T}$ at $m_\pi \gg T$.

(ii) The chiral condensate
$\la \bar{u} u\ra_T$ changes as $\la\bar{u}u\ra_0
(1-\zeta/8)$.
Its $T$-dependence was calculated by the chiral perturbation theory
including up to $O(T^6)$ effects\,\cite{GL2},
which gives the same coefficient as above for the $O(T^2)$ effect.
In the temperature range $T\leq 160$ MeV, both calculations
agree well within 5 \%.

(iii) By using the
QCD trace anomaly, $T$-dependence of $\la{\alpha_s\over \pi}G^2\ra_T$
can be estimated\,\cite{HKL}.
It gives a negligible change of the condensate at finite
$T$ (0.5\% at $T=200$ MeV).
We thus ignored its $T$-dependence in (\ref{eq210}).

(iv) The four-quark condensate in the nucleon channel turned out
to be $T$-independent, which is quite different from the behavior
of the square of the chiral order parameter.
Here we used the vacuum saturation assumption, i.e.,
$\la (\bar{u}\Gamma u)^2 \ra \rightarrow \la \bar{u}u \ra^2$ {\it after}
applying the soft pion theorem, as is usually adopted in the vacuum QSR.
The calculation of the four-quark condensates
is somewhat tedious, so we shall present a nonfactorized form
of the four-quark condensates for the octet baryons
in the Appendix.
There one sees that the $T$-dependence of the four-quark condensates
is different in different channels.
For example, the four-quark condensate in the nucleon channel is
$T$-independent, while it is proportional to $1-\zeta/6$
in the $\Pi_1$ structure in the
$\Sigma$-channel (See also (\ref{eq211})--(\ref{eq214}) in the Appendix).

\subsection{Spectral function}

In the pion gas, the spectral function for the nucleon current acquires
a contribution from the scattering with the thermal pions:
$\pi + N \rightarrow N, \Delta$ etc.
The effect of these scatterings
has to be taken into account in the spectral function before
the change of the condensates is ascribed to the shift of the pole position.
For the purpose of identifying these contributions,
we introduce the expression for the spectral
function\cite{AGD}:
\beq
\rho_{\alpha\beta}(\omega,\qvec,T)&=&{1\over \pi}{\rm Im}\Pi^R_{\alpha\beta}
(\omega,\qvec,T)\nonumber\\
&=&{1\over Z}(2\pi)^3 \sum_{n,m} \la n|\eta_\alpha(0)|m\ra
\la m|\bar{\eta}_\beta(0)|n\ra \nonumber\\
& & \qquad\qquad \times(e^{-\varepsilon_n/T} + e^{-\varepsilon_m/T})
\delta(\omega-\omega_{mn})\delta^{(3)}(\qvec-\pvec_{mn}),
\label{eq217}
\eeq
where the states $|m\ra$ and $|n\ra$ have the four momentum
$(\varepsilon_m, \pvec_m)$ and $(\varepsilon_n, \pvec_n)$, respectively,
and
$ \omega_{mn}=\varepsilon_m - \varepsilon_n$, $\pvec_{mn}=
\pvec_{m}-\pvec_{n}$.
If we put $|n\ra=|0\ra$, $|m\ra=|N(p)\ra$ (
$|m\ra=|0\ra$, $|n\ra=|{\bar N}(p)\ra$) in (\ref{eq217}),
this is the contribution
from the nucleon (anti-nucleon) at $T=0$.
By introducing the nucleon and the anti-nucleon spinor by the relation
\beq
\la 0|\eta_{\alpha}(0)|N(p)\ra = \lambda_N u_\alpha(p),\qquad\qquad
\la 0|{\bar \eta_{\alpha}}(0)|{\bar N}(p)\ra = \lambda_N
\bar{v}_\alpha(p)
\label{eq218}
\eeq
with the normalization ${\bar u}(p)u(p) = 2M_N$ and
${\bar v}(p)v(p) = -2M_N$, we get for this contribution
\beq
\rho(\omega,\qvec) = {\lambda_N^2\over{2p_0}}
(\qslash +M_N)\left(\delta(\omega-p^0)-
\delta(\omega+p^0)
\right)
\label{eq219}
\eeq
with $p^0=\sqrt{\qvec^2+M_N^2}$.
Using the Borel sum rule method, we can study the $T$-dependence of
the mass $M_N(T)$, of the pole residue $\lambda_N(T)$, and of the continuum
threshold $S_0(T)$.
However, it is difficult to incorporate the effects of the width and the
scattering contribution induced in the pion gas.
Thus we shall first list up the effects which should not be associated with
the change of the
above three resonance parameters. Then we put these additional structures
at $T\neq 0$ in the spectral function when we carry out the Borel sum rule.

(i) $\pi + N \rightarrow N$ ($\pi + \bar{N} \rightarrow \bar{N}$)
contribution; $|n\ra=|\pi(k)\ra$, $|m\ra=|N(p)\ra$
($|n\ra=|\bar{N}(p)\ra$, $|m\ra=|\pi(k)\ra$):
By taking into account the Bose symmetrization among pions which
equally fill both $|n\ra$ and $|m\ra$, but do not interact with the nucleon
current, we arrive at
\beq
\rho^{\pi+N\rightarrow N}(q,T) &=&
(2\pi)^3\int {d^3k \over (2\pi)^3 2k^0 }n_B(k^0/T)
\int {d^3p \over (2\pi)^3 2p^0 } \nonumber\\[10pt]
& & \times
\left[
\sum_{{\rm spin}=\pm 1/2}\sum_a \la \pi^a(k)|\eta_\alpha(0)|N(p)\ra
\la N(p)|\bar{\eta}_\beta(0)|\pi^a(k)\ra \right. \nonumber\\[10pt]
& & \left. \times
\left( \delta(\omega - p^0 + k^0)
\delta^{(3)}(\qvec - \pvec + \kvec ) +
\delta(\omega - p^0 - k^0)
\delta^{(3)}(\qvec - \pvec - \kvec) \right) \right. \nonumber\\[10pt]
& & \left. +
\sum_{{\rm spin}=\pm 1/2}\sum_a \la \bar{N}(p)|\eta_\alpha(0)|\pi^a(k)\ra
\la \pi^a(k)|\bar{\eta}_\beta(0)|\bar{N}(p)\ra \right. \nonumber\\[10pt]
& & \left. \times
\left( \delta(\omega + p^0 + k^0)
\delta^{(3)}(\qvec + \pvec + \kvec) +
\delta(\omega + p^0 - k^0)
\delta^{(3)}(\qvec + \pvec - \kvec) \right)
\right]. \nonumber\\
\label{eq220}
\eeq
Here we discarded the thermal factor for the nucleon
(Fermi distribution function),
since $1/(e^{M_N/T}-1)\simeq 0$ at $T\leq 200$ MeV.
We have to include two kinds of contribution to the matrix element
$\la \pi^a(k)|\eta_\alpha(0)|N(p)\ra$\,\cite{Ad}.

(a) Direct coupling of $\eta^N$ to the pion (Fig.\,1):  This contribution
can be calculated by applying the soft pion theorem:
\beq
\la\pi^a(k)|\eta^{\rm p}_\alpha(0)|N(p)\ra &=& {-i\over f_\pi }
\la 0|\left[{\cal F}_5^a, \eta^{\rm p}_\alpha(0) \right]|N(p)\ra,
\label{eq221a}
\\[10pt]
\left[{\cal F}_5^a, \eta^{\rm p}_\alpha(0) \right]
&=& \left({\tau^a \over 2}\right)_{22}\gamma_5 \eta^{\rm p} -
 \left({\tau^a \over 2}\right)_{12}\gamma_5 \eta^{\rm n},
\label{eq221b}
\eeq
where $\eta^{\rm p}$ is the proton current defined in (\ref{eq1}) and
$\eta^{\rm n} = -\epsilon^{abc}d^a(x)C\gamma_\mu d^b(x)
\gamma_5\gamma_\mu u^c(x) $ is the neutron current.
Using (\ref{eq221a}) and (\ref{eq221b}) in (\ref{eq220})
and putting $\qvec={\bf 0}$, we arrive at
\footnote{Here $\pvec$ has $\kvec$ or $-\kvec$ corresponding to two
$\delta^{(3)}$-functions in (\ref{eq220}).}
\beq
\rho^{\pi+N\rightarrow N}_{(0)}(q,T)
&=& \left( {-3\lambda_N^2 \over 4 f_\pi^2 }\right)
\int {d^3k \over (2\pi)^3 2k^02p^0 }n_B(k^0/T)\times \nonumber\\[10pt]
& & \left[
\gamma_5 (\pslash+M_N)\gamma_5
\left( \delta(\omega - p^0 + k^0) + \delta(\omega - p^0 - k^0) \right)
\right.\nonumber\\[10pt]
& & +\left.\gamma_5 (\pslash-M_N)\gamma_5
\left( \delta(\omega + p^0 + k^0) + \delta(\omega + p^0 - k^0) \right)
\right].
\label{eq222}
\eeq
Since we used the soft pion theorem in calculating the matrix element
in (\ref{eq221a}),
it suffices to consider the soft pion limit
$k=(k^0,\kvec) \rightarrow 0$ in (\ref{eq222}).
Then (\ref{eq222}) becomes
\beq
\rho^{\pi+N\rightarrow N}_{(0)}(q,T) =
\left({\lambda_N^2\zeta \over 32 M_N }\right)
\left( \delta(\omega-M_N) - \delta(\omega+M_N) \right)
\left(\qslash -M_N\right).
\label{eq225}
\eeq
Equation (\ref{eq222}) was derived by Eletsky\,\cite{El} taking the
imaginary part of the corresponding retarded correlator.\footnote
{The result given in eq.\,(11) of \cite{El} is two times larger than
the one given in
(\ref{eq222}) by mistake.  I thank V.\,L.\,Eletsky for correspondence
to clarify this point.}
There it was shown
that ({\ref{eq222}) has a localized structure around
the nucleon and anti-nucleon poles, which can be well approximated by
(\ref{eq225}).

(b) Coupling of $\eta^N$ to the nucleon which interacts with $\pi$ (Fig.\,2):
This contribution can be calculated by using the vertex
\beq
\la \pi(k)| \eta(0)|N(p)\ra = \lambda_N\la \pi(k)| \psi^N(0)|N(p)\ra
={\lambda_N g_{_{\pi NN}}\over 2M_N } {1 \over \pslash-\kslash -M_N }
\kslash\gamma_5 u(p),
\label{eq226}
\eeq
where we assumed the $\pi-N-N$ interaction lagrangian as
\beq
{\cal L}_{\pi NN} = {g_{_{\pi NN}} \over 2M_N} \bar{N}\gamma_5\gamma_\mu
\vec{\tau}N \dmu\vec{\pi}.
\label{eq227}
\eeq
Inserting (\ref{eq226}) into (\ref{eq220}),
one obtains in the $\qvec={\bf 0}$ limit:
\beq
\rho^{\pi+N\rightarrow N}_{(1)}(q,T) &=&
\int {d^3k \over (2\pi)^3 2k^0 2p^0 }n_B(k^0/T)
\,3\left({\lambda_N g_{_{\pi NN}} \over 2 M_N }\right)^2
{ 1 \over \left(q^2-M_N\right)^2 } \nonumber\\[10pt]
& & \times \left[
\left(\qslash+M_N\right)\kslash\gamma_5\left(\pslash+M_N\right)
\kslash\gamma_5\left(\qslash+M_N\right) \right.\nonumber\\[5pt]
& & \left.\qquad\qquad \times
\left( \delta(\omega - p^0 + k^0) + \delta(\omega - p^0 - k^0) \right)
\right.
\nonumber\\[10pt]
& & \left. +
\left(-\qslash+M_N\right)\kslash\gamma_5\left(\pslash-M_N\right)
\kslash\gamma_5\left(-\qslash+M_N\right) \right.\nonumber\\[5pt]
& & \left. \qquad\qquad \times
\left( \delta(\omega + p^0 + k^0) + \delta(\omega + p^0 - k^0) \right)
\right]
\label{eq228}\\[15pt]
&=& \left({{g_{_A}}^2\lambda_N^2\zeta \over 32 M_N }\right)
\left( \delta(\omega-M_N) - \delta(\omega+M_N) \right)
\left( \qslash +M_N \right).
\label{eq231}
\eeq
In (\ref{eq231}),
we used the Goldberger-Treiman relation $g_{_{\pi NN}}/M_N=g_{_A}/f_{\pi}$.
Note that in (\ref{eq225})
the two structures proportional to ${\bf 1}$ and $\qslash$
have opposite signs while they have the same sign in (\ref{eq231}).

One can easily check that the crossing term between (a) and (b) disappears.

In the above (a) and (b), we have obtained the $\pi+N\rightarrow N$ scattering
contribution to the spectral function at $\qvec={\bf 0}$ in the form:
\beq
\rho^{\pi+N\rightarrow N}(\omega,T) &=&
\rho^{\pi+N\rightarrow N}_{(0)}(q,T)
+\rho^{\pi+N\rightarrow N}_{(1)}(q,T)\nonumber\\[10pt]
&=&\left({\lambda_N^2\zeta \over 32 M_N }\right)
\left( \delta(\omega-M_N) - \delta(\omega+M_N) \right)
\left\{\left(1+g_{_A}^2\right) \qslash - \left(1-g_{_A}^2\right)M_N \right\}.
\nonumber\\
\label{eq231b}
\eeq
We note that up to $O(T^2)$ the use of the nucleon mass
and the pole residue of $T=0$
in (\ref{eq231b})
is consistent with our treatment of the OPE side because of
the presence of the factor $\zeta$ in (\ref{eq231b}).
(From a physical ground, one might wish to replace them by those
at $T\neq 0$.  But these two procedures
cause only an $O(T^4)$ difference in the final result.)

In (\ref{eq231b}), the $\pi+N \rightarrow N$ scattering term eventually
becomes an effective delta function at the pole position of $T=0$.
However, it contributes differently to the ${\bf 1}$ and $\qslash$
structures of the correlators.  Therefore its effect
must be taken into account in the spectral function when we make use of
the Borel sum rule method.  Otherwise, an erroneous mass shift of the
nucleon would occur.

(ii) $\pi+N \rightarrow \Delta$ contribution:  Using the lowest
order piece of the chiral invariant $\pi-N-\Delta$ interaction lagrangian
${\cal L}_{\pi N \Delta} \sim g_{_{\pi N \Delta}}
\bar{\Delta}_\mu\gamma_5 N \dmu\pi$\,\cite{Pe},
we can calculate the vertex:
\beq
\la\pi(k)|\eta(0)|\Delta(p)\ra \sim {\lambda_N g_{_{\pi N \Delta}}
\over \pslash -\kslash - M_N} k^\mu
\Delta_\mu(p),
\label{eq232}
\eeq
where $\Delta_\mu(p)$ is the Rarita-Shwinger spinor for $\Delta$.
The sum over spin for $\Delta$ gives the projection operator\,\cite{Pe}
\beq
\sum_{spin} \Delta_\mu (p)\Delta_\nu (p) = {\cal P}_{\mu\nu}
=\left[ g_{\mu\nu} -
{2 \over 3 M_\Delta^2}p_\mu p_\nu - {1 \over 3}\gamma_\mu \gamma_\nu
- {1 \over 3M_\Delta}(p_\mu\gamma_\nu - p_\nu\gamma_\mu)\right]
(\pslash + M_\Delta).
\label{eq233}
\eeq
In the soft pion limit,
the $\pi + N \rightarrow \Delta$
scattering term appears at $\omega \sim M_\Delta$
where the nucleon propagator in (\ref{eq232}) becomes proportional to
$1/(M_\Delta^2 - M_N^2)$ and the vertex contribution is
$k^\mu k^\nu {\cal P}_{\mu\nu} \sim k^2$.  Therefore
the $\pi + N \rightarrow \Delta$ scattering contribution becomes $O(T^4)$ and
we shall discard it.

One can easily repeat the same steps as above (i) (a) (b) and (ii)
for the other octet baryons.  By the same reason as (ii), the
effect of the transition to
the decuplet baryons ($\pi+\Sigma \rightarrow \Sigma^*(1385)$,
$\pi+\Xi\rightarrow\Xi^*(1530)$) is $O(T^4)$.  Even among octet
baryons, (i)(b) type scattering between $\Lambda$ and $\Sigma$ is
$O(T^4)$ because of their mass difference.

\subsection{Borel sum rule}
We are now ready to perform the Borel sum rule analysis for the octet baryons.
We assumed that the finite-$T$ medium
is the dilute pion gas which has zero chemical potential, and thus
the charge conjugation symmetry is preserved, i.e., there appears
no splitting between the baryon and anti-baryon poles.\footnote{
This is in contrast to the system with a finite baryon number
such as the nuclear matter.  In order to organize
sum rules for baryons in such a system, the dispersion relation
needs some modification.  See \cite{KM} for the detail.}
Therefore the spectral function for the nucleon reads at $\qvec={\bf 0}$
\beq
\rho^N(u,T) &=& \lambda_N^2(T) \delta(u^2 -M_N^2(T))(\qslash + M_N(T))
{\rm sign}(u)
+\rho^{\pi+N\rightarrow N}(u,T)\nonumber\\[5pt]
& &\qquad\qquad + (\rm continuum\ by\ step\ function),
\label{eq302}
\eeq
where $\rho^{\pi+N\rightarrow N}$
is defined in (\ref{eq231b}).
Corresponding to (\ref{eq202}), we decompose the spectral
function as $\rho^N(q) = \rho_1(q)
+ \qslash \rho_2(q)$.
Then both $\rho_1$ and $\rho_2$
satisfy the relation $\rho_i(-\omega,\qvec={\bf 0})
=-\rho_i(\omega,\qvec={\bf 0})$ $(i=1,2)$ as was
obtained in the previous subsection.
In the deep Euclidean region,
the dispersion relation (\ref{eq6}) can be written as
\beq
\Pi_i^R(\omega^2=-Q^2,\qvec={\bf 0},T)=\Pi_i(Q^2,T)
= \int_0^{\infty}du
{ 2u\rho_i(u,\qvec={\bf 0}) \over u^2 + Q^2 }.\ \ \ (i=1,\ 2)
\label{eq301}
\eeq
Applying the Borel transform
\beq
\Pi_i(M^2,T)&\equiv& \hat{B}_M \Pi_i(Q^2,T) \nonumber\\[10pt]
&\equiv&
\lim_{ \left( \begin{array}{c}
Q^2, n \rightarrow\infty\\
Q^2/n=M^2:{\rm fixed} \end{array} \right)  }
{ 1 \over (n-1)!} (Q^2)^n\left( {-d \over dQ^2 } \right)^n
\Pi_i(Q^2,T)
\label{eq303}\\[10pt]
&=& \int_0^\infty 2udu\,e^{-u^2/M^2}\rho_i(u,T)
\label{eq304}
\eeq\\
to (\ref{eq209}) and (\ref{eq210})
with the spectral function (\ref{eq302}),
we get the following relation:\\
\beq
2a\left(1-{\zeta\over 8}\right)M^4
= \tilde{\lambda}_N^2(T)M_N(T)e^{-M_N^2(T)/M^2}
-{\tilde{\lambda}_N^2(0)(1-{g_{_A}}^2)
M_N(0)\zeta\over 16}e^{-M_N^2(0)/M^2},
\label{eq305}\\[18pt]
M^6+M^2b+{4\over3}a^2 = \tilde{\lambda}_N^2(T)e^{-M_N^2(T)/M^2}
+{\tilde{\lambda}_N^2(0)(1+{g_{_A}}^2)\zeta\over 16}e^{-M_N^2(0)/M^2},
\label{eq306}\\
\nonumber\eeq
where
\beq
a &=& -(2\pi)^2\la\bar{u}u\ra,
\label{eq307}\\[10pt]
b &=& \pi^2\la {\alpha_s \over \pi} G^2 \ra,\label{eq308}
\\[10pt]
\tilde{\lambda}^2_N(T) &=& 2(2\pi)^4\lambda^2_N(T).
\label{eq310}
\eeq
{}From (\ref{eq305}) and (\ref{eq306})
one gets the expression for the nucleon mass:
\beq
M_N(T) = { 2a\left(1-{\zeta\over 8}\right)M^4 +
\left( \tilde{\lambda}^2_N(0)(1-{g_{_A}}^2) M_N(0)\zeta/16\right)
e^{-M_N^2(0)/M^2}
\over
M^6 + M^2b + {4\over 3}a^2 -
\left( \tilde{\lambda}^2_N(0)(1+{g_{_A}}^2) \zeta/16\right)
e^{-M_N^2(0)/M^2}
}.
\label{eq311}
\eeq
In (\ref{eq311}),
we omitted the correction due to the continuum contribution for brevity.
The formula which include the correction
is given by the following replacement:
\beq
M^6 &\rightarrow & M^6\left\{ 1 -\left( 1 +  {S_0 \over M^2} +
{S_0^2 \over 2 M^4}\right) e^{-S_0/M^2} \right\},\nonumber\\
M^4 &\rightarrow & M^4\left\{ 1 -\left( 1 +  {S_0 \over M^2}
\right) e^{-S_0/M^2} \right\},\nonumber\\
M^2 &\rightarrow & M^2\left( 1 - e^{-S_0/M^2} \right).
\label{eq315}
\eeq
An important consequence from (\ref{eq311})
is that $M_N(T)$ is completely $T$-independent.
In fact, if we replace $\tilde{\lambda}_N^2(0)$ in the numerator of
(\ref{eq311}) by the one obtained from (\ref{eq305}) by setting $T=0$,
and replace $\tilde{\lambda}_N^2(0)$
in the denominator of (\ref{eq311}) by the one obtained
from (\ref{eq306}), we can easily see that the $T$-dependence disappears
from $M_N(T)$.
If we did not include the scattering term, the nucleon mass
would behave as $M_N(T)=M_N(0)(1-\zeta/8)$ as can be seen from (\ref{eq311}).
This means that the $T$-dependence of the condensates caused through the
interaction with the thermal pions is completely compensated by the
$\pi-N$ scattering term and the pole
position does not move at least to order $O(T^2)$.
This is consistent with
the statement of Leutwyler and Smilga based on the chiral
lagrangian\,\cite{LS}:
If we calculate the self
energy of the nucleon using the $\pi NN$ effective lagrangian (\ref{eq227}),
we can easily check that the real part of the self energy is zero at
the $T=0$ pole position.  This is a direct consequence of the Adler's
consistency condition required for the use of PCAC\,\cite{Ad}
(in another word, (\ref{eq227}) has a derivative coupling).

The pole residue changes as $\tilde{\lambda}^2_N(T)=\tilde{\lambda}^2_N(0)
(1-(1+g_{_A}^2)\zeta/16)$ as is seen from (\ref{eq305}) and (\ref{eq306});
the same result obtained by \cite{LS} in the chiral lagrangian approach.

{}From the above demonstration, it is clear that we have to take into account
the new structure in the spectral function consistently
with the $T$-dependence in the OPE side of the correlator.
Otherwise
the usual procedure in the sum rule leads to an artificial
change of the resonance parameters.

In \cite{AZ}, the nucleon mass was calculated by a finite-$T$ QCD sum rule
method.
The authors found a dropping nucleon mass
even in the low temperature region as $\la\bar{u}u\ra_T$
decreased at finite-$T$.
They did not take into account the $\pi+N\rightarrow N$ scattering effect
and
assumed that the $T$-dependence of the four-quark condensate
is the same as $(\la\bar{u}u\ra_T)^2$.
Although their calculation was not based on the pion gas approximation,
the present consideration shows it is crucial to treat
both the OPE side and the phenomenological side consistently.
Correct treatment of the $T$-dependence of all the condensates
is also required.

\section{Analysis of Correlators using PCAC}
\setcounter{equation}{0}
\renewcommand{\theequation}{\arabic{section}.\arabic{equation}}
In this section we shall examine the octet
baryon correlators starting from the pion gas
approximation without using OPE:
\beq
\Pi(q,T)
\simeq \Pi(q,0) + i\int\,d^4x\, e^{iqx}\int{d^3k \over
(2\pi)^3 2k^0} n_B(k^0/T)
\la \pi^a(k)|T\left(\eta(x) \bar{\eta}(0)\right)|\pi^a(k)\ra.
\label{eq401}
\eeq
Applying the LSZ reduction formula for the pion, then using
the PCAC relation $\d^\mu A_\mu^a(x) = f_\pi m_\pi^2\phi^a(x)$
for the pion field $\phi^a(x)$
and taking the soft pion limit, one arrives at
\beq
\Pi(q,T) &\simeq& \Pi(q,0)
 - {i\delta^{ab}\zeta \over 24}
\int\,d^4x\, e^{iqx} \times
\nonumber\\[10pt]
& &\ \left\{
\la T\left( \left[{\cal F}_5^a(x^0),
\left[{\cal F}_5^b(x^0), \eta(x)\right]\right]\bar{\eta}(0)
\right) \ra
+\la T\left( \eta(x)\left[{\cal F}_5^a(0),
\left[{\cal F}_5^b(0),\bar{\eta}(0)\right]\right]
 \right) \ra
\right.\nonumber\\[10pt]
& &\left.
+\la T\left(\left[{\cal F}_5^a(x^0), \eta(x)\right]
\left[{\cal F}_5^b(0),\bar{\eta}(0)\right]
 \right) \ra
+\la T\left(\left[{\cal F}_5^b(x^0), \eta(x)\right]
\left[{\cal F}_5^a(0),\bar{\eta}(0)\right]
 \right) \ra
\right\}\nonumber\\
& & +\cdot\cdot\cdot,
\label{eq402}
\eeq
where $+\cdot\cdot\cdot$ denotes the terms associated with the axial charges
carried by the octet baryons (terms with $g_{_A}$
in the previous section), which becomes $O(T^4)$ or higher except at
the pole position of $T=0$.\footnote{In the sum rule, we needed to
integrate over the spectral functions, which is why the $O(T^2)$ contribution
appeared.}
Utilizing the formula (\ref{eq221b})
together with
\beq
\delta^{ab}\left[{\cal F}^a_5,\left[{\cal F}^b_5,\eta^N\right]\right]
={3\over 4} \eta^{N},\label{eq403}
\eeq
we obtain the following expression for the nucleon correlator at $T\neq 0$:
\beq
\Pi^N(q,T) &=& \left( 1 - {\zeta \over 16 } \right) \Pi^N(q,0)
-{\zeta \over 16 } \gamma_5 \Pi^N(q,0) \gamma_5 +\cdot\cdot\cdot.
\label{eq407}
\eeq
The second term of the r.h.s. of (\ref{eq407})
corresponds to
the $\pi+N \rightarrow N$
scattering term in the sum rule approach in the previous section
and the first term of the r.h.s. of
(\ref{eq407})
is the modification of the residue of the
nucleon current.  The above analysis of the current
simply tells us that
the nucleon correlator at $T\neq 0$ can be written as
a superposition of the same correlator at $T=0$ with $T$-dependent
coefficients and
there is no
$O(T^2)$ shift of the pole position, which is consistent with
the observation of \cite{LS}.
(As was shown in \cite{LS} using (\ref{eq227}),
the contribution to the real part of the self-energy
from $+\cdot\cdot\cdot$ in (\ref{eq407}) becomes zero at the pole position
but induces an $O(T^2)$ wave function renormalization.)

Equation (\ref{eq407}) reads
$\Pi^N_1(q,T)=\Pi^N_1(q,0)(1-\zeta/8)+\cdot\cdot\cdot$ and
$\Pi^N_2(q,T)=\Pi^N_2(q,0)+\cdot\cdot\cdot$.  The $T$-dependence of
these two equations is the same as the OPE given in
(\ref{eq209}) and (\ref{eq210}).
This is the reason we observed
no mass shift in the Borel sum rule.
We remind the readers once again that the factorization of the
four-quark condensate in the medium level
$\la (\bar{q}\Gamma q)^2\ra_T \rightarrow \la\bar{q}q\ra_T^2$ is
not justified.  If we adopted this procedure in section 2, the $T$-dependence
of the OPE side of the correlator would be different from (\ref{eq407}).

It is easy to extend the above analysis to other octet baryon correlators.
For this purpose we need the following commutators:
\beq
\left[{\cal F}^a_5, \eta^\Lambda \right] &=& \sqrt{2 \over 3} \gamma_5
\left[ -\left({\tau^a \over 2}\right)_{21} \eta^{\Sigma^+}
       +\left({\tau^a \over 2}\right)_{12} \eta^{\Sigma^-}
+ {1 \over \sqrt{2} }\left(
\left({\tau^a \over 2}\right)_{11} -
\left({\tau^a \over 2}\right)_{22} \right) \eta^{\Sigma^0} \right],\
\label{eq234}\\[10pt]
\left[{\cal F}^a_5, \eta^{\Sigma^+} \right]&=&{\tau^a_{12} \over \sqrt{2} }
\gamma_5 \eta^{\Lambda'},
\label{eq242}\\[10pt]
\left[{\cal F}^a_5, \eta^{\Xi^0} \right] &=&
\left({\tau^a \over 2}\right)_{11}\gamma_5 \eta^{\Xi^0}
       +\left({\tau^a \over 2}\right)_{12}\gamma_5 \eta^{\Xi^-},
\label{eq235}
\eeq
\beq
\delta^{ab}\left[{\cal F}^a_5,\left[{\cal F}^b_5,\eta^\Lambda\right]\right]
&=&-\sqrt{3} \eta^{\Lambda'},\label{eq404}\\[10pt]
\delta^{ab}\left[{\cal F}^a_5,\left[{\cal F}^b_5,\eta^\Sigma\right]\right]
&=& \eta^{\Sigma},\label{eq405}\\[10pt]
\delta^{ab}\left[{\cal F}^a_5,\left[{\cal F}^b_5,\eta^\Xi\right]\right]
&=&{3\over 4} \eta^{\Xi},\label{eq406}
\eeq
where we introduced a new current $\eta^{\Lambda'}$ defined by
\beq
\eta^{\Lambda'}= \sqrt{2}\epsilon^{abc}u^aC\gamma_5\gamma_\mu d^b
\gamma_\mu s^c,
\label{eq243}
\eeq
in ({\ref{eq242}) and (\ref{eq404}).
By using these relations in (\ref{eq402}),
we obtain the following relations among the correlators:
\beq
\Pi^\Lambda(q,T) &=&  \Pi^\Lambda(q,0) +
{\sqrt{3}\zeta \over 24 } \Pi^{\Lambda\Lambda'}(q,0)
-{\zeta \over 12 } \gamma_5 \Pi^{\Sigma}(q,0) \gamma_5 +\cdot\cdot\cdot,
\label{eq408}\\[15pt]
\Pi^\Sigma(q,T) &=& \left( 1 - {\zeta\over 12 } \right) \Pi^\Sigma(q,0)
-{\zeta\over 12 } \gamma_5 \Pi^{\Lambda'}(q,0) \gamma_5 +\cdot\cdot\cdot,
\label{eq409}\\[15pt]
\Pi^\Xi(q,T) &=& \left( 1 - {\zeta\over 16 } \right) \Pi^\Xi(q,0)
-{\zeta \over 16 } \gamma_5 \Pi^\Xi(q,0) \gamma_5 +\cdot\cdot\cdot,
\label{eq410}
\eeq
with
\beq
\Pi_{\alpha\beta}^{\Lambda\Lambda'}(q,0) =
i\int\,d^4x\,e^{iqx}\la T\left(
\eta_{\alpha}^\Lambda(x)
\bar{\eta}_{\beta}^{\Lambda'}(0)+
\eta_{\alpha}^{\Lambda'}(x)
\bar{\eta}_{\beta}^{\Lambda}(0)
\right)\ra.
\label{eq411}
\eeq
The current $\eta^{\Lambda'}$ is anti-symmetric under the exchange between
$u$ and $d$ quarks and thus $\eta^{\Lambda'}$
(or $\gamma_5\eta^{\Lambda'}$) should have some overlap with isoscalar
strangeness=--1 baryons such as $\Lambda(1115)$, $\Lambda^*(1405)$
($J^P=1/2^-$) as well
as the $\pi-\Sigma$ continuum contribution.  As is shown in \cite{EPT},
there are 5 independent interpolating fields without derivatives for
$\Lambda$.  One can see by the Fierz rearrangement
between $s$ and $u$ (or $d$) that $\eta^{\Lambda'}$
consists of $\eta^{\Lambda}$ and those others.
We tried to identify its structure by the vacuum QSR including the
operators up to dimension-6, but it does not seem to have a
dominant pole contribution.  In any case, (\ref{eq409})
tells us that the finite-$T$ $\Sigma$-correlator can be written
as the modification of the residue and the mixing with $\Lambda'$
correlator. In principle, the second term of (\ref{eq408})
also describes the modification of the residue.
As for $\Xi$, the situation is completely
parallel with the nucleon.
The $T$-dependence of the OPE expressions (\ref{eq214}) and
(\ref{eq215}) is consistent with
(\ref{eq410}) as they should.

\section{Summary and Outlook}
\setcounter{equation}{0}
\renewcommand{\theequation}{\arabic{section}.\arabic{equation}}
In this paper we have presented an explicit demonstration that
the $O(T^2)$ dependence of the condensates which appear in the OPE
of the octet baryon correlators is totally absorbed by
the scattering terms $\pi+B\rightarrow B'$ and the
modifications of the pole residues, in the framework of the QCD
sum rules.  This result is consistent with the statement by Leutwyler and
Smilga\,\cite{LS}.
The result stems from the fact that
the baryon correlators in the thermal pion gas can be
written as a superposition
of the correlators of $T=0$ with $T$-dependent coefficients up to $O(T^2)$.
The procedure for achieving consistency with
this relation is somewhat intricate in the QSR. Therefore
one has to pay particular attention to
the consistency between the assumption
made to estimate the $T$-dependence of the condensates
and the new structure appearing in the phenomenological
spectral function.

In the chiral lagrangian approach, the $O(T^2)$ mass shift of baryons
is caused by tadpole interactions such as
$m_sKK\bar{B}B/f_\pi^2$ ($K$ is the kaon field)\,\cite{BSW}.
In the $m_u=m_d=0$, $m_s\neq 0$ limit studied in this work,
the kaon or $\eta$ field
always accompanies the tadpole contribution
as in the case of the above interaction.
Thus without including kaons or $\eta$
in the heat bath, baryons do not receive a $O(T^2)$ mass shift.
However, the presence of those massive
excitations is suppressed as $\sim e^{-m_K/T}$.
For example, the above
term in the effective lagrangian causes the mass shift of the order of
$m_sT^2/(24f_\pi^2)B_1(m_K/T) \sim 1$ MeV at $T=150$ MeV.
If we included the kaons and $\eta$'s in the heat bath together
with the nonzero strange-quark mass, the $T$-dependence of the OPE
for the hyperons shown in (\ref{eq211})--(\ref{eq215})
would be different from (\ref{eq408})--(\ref{eq410}).
This would
lead to the $O(T^2)$ mass shift of the octet baryons,
although it should be tiny because of $B_1(m_K/T)$ ($\sim 0.06$ at
$T=150$ MeV).

To go beyond $O(T^2)$, one needs more information on the pion structure
functions (twist-4 effects), $O(T^4)$ or higher $T$-dependence of
the scalar condensates and the analysis of the structure $\Pi_3$
in (\ref{eq202})
as well as more involved treatment for the phenomenological side
(such as octet $\rightarrow$ decuplet transitions).
These issues are beyond the scope of the present study.
I hope the lesson we learned through the demonstration in this work
will be useful
for more advanced studies on these higher order effects.

\vspace{1 cm}
\cl{\large{\bf Acknowledgement}}

I would like to thank T.\,Hatsuda for illuminating discussion
and useful comments on the manuscript.  I'm also grateful to
P.\,Danielewicz for useful discussions and reading the manuscript.
Conversations with S.\,H.\,Lee,
T.\,Matsui, S.\,Nagamiya and K.\,Yazaki are also acknowledged.
This work is supported in part by the US National Science Foundation
under grant PHY-9017077.

\newpage
\cl{\large{\bf Appendix}}
\setcounter{equation}{0}
\renewcommand{\theequation}{A.\arabic{equation}}
Here we will present the nonfactorized form of the
four-quark operators
which appear in the OPE for the octet baryon currents.
Nonscalar four-quark operators are associated with the
twist-4 contribution in the pion structure function.
We ignore them since they become
$O(T^4)$ effect in the QCD sum rules at $T\neq 0$.
We decompose the contribution of the four-quark condensate
into the two pieces corresponding to the two structures of the correlators:
\beq
\Pi^{N,\Lambda,\Sigma,\Xi}_{\rm 4-quark}(q) =
\tilde{\Pi}^{N,\Lambda,\Sigma,\Xi}_1
+ \tilde{\Pi}^{N,\Lambda,\Sigma,\Xi}_2 \qslash.
\label{a0}
\eeq
(i) {\it N}:\\
\beq
\tilde{\Pi}_1^N(q)&=&0,
\label{a1}\\[10pt]
\tilde{\Pi}_2^N(q)&=& {-1 \over q^2}
\left[
\left(1-{1\over N_c}\right)
\left\{
{5\over 2}\la\bar{u}\gamma_\mu u\bar{d}\gamma_\mu d \ra
+{3\over 2}\la\bar{u}\gamma_\mu\gamma_5 u\bar{d}\gamma_\mu\gamma_5 d \ra
\right\}
\right. \nonumber\\[10pt]
& & \left. \qquad\qquad
-{1\over 2}
\left\{
{5\over 2}\la\bar{u}\gamma_\mu \lambda^a u
\bar{d}\gamma_\mu \lambda^a d \ra
+{3\over 2}\la\bar{u}\gamma_\mu \gamma_5 \lambda^a u
\bar{d}\gamma_\mu\gamma_5 \lambda^a d \ra
\right\}
\right. \nonumber\\[10pt]
& & \left.
+ \left(1-{1\over N_c}\right)
\left\{
{1\over 2}\la(\bar{u} u)^2 \ra - {1\over 2}\la(\bar{u}\gamma_5 u)^2 \ra
-{1\over 4}\la(\bar{u}\gamma_\mu\gamma_5 u)^2 \ra
+{1\over 4}\la(\bar{u}\gamma_\mu u)^2 \ra
\right\}
\right. \nonumber\\[10pt]
& & \left.
-{1\over 2}\left\{
{1\over 2}\la(\bar{u}\lambda^a u)^2 \ra
- {1\over 2}\la(\bar{u}\gamma_5 \lambda^a u)^2 \ra
-{1\over 4}\la(\bar{u}\gamma_\mu\gamma_5 \lambda^a u)^2 \ra
+{1\over 4}\la(\bar{u}\gamma_\mu \lambda^a u)^2 \ra
\right\}
\right],\nonumber\\
\label{a2}
\eeq
where $\lambda^a$ is the SU(3) color matrix and
we explicitly kept the $N_c (=3)$ dependence.
As is seen from (\ref{a2}), the four-quark operators
always appear in the form of
\beq
\left(1-{1\over N_c}\right) \{ \bar{q}\Gamma q\bar{q'}\Gamma q'
+\cdot\cdot \}
-{1 \over 2} \{ \bar{q}\Gamma\lambda^a
 q\bar{q'}\Gamma\lambda^a q'+\cdot\cdot \}.\nonumber
\eeq
We will henceforth use
the abbreviation $-{1\over 2}\{ {\rm with}\
\lambda^a \}$ to denote the second contribution.

\noindent
(ii) $\Lambda$:\\
\beq
\tilde{\Pi}^\Lambda_1(q) &=& {-m_s\over 3 q^2}
\left[
\left( 1 -{1\over N_c } \right)
\left\{
4\la \bar{u} u \bar{d} d \ra
- \la \bar{u}\sigma_{\mu\nu} u \bar{d}\sigma_{\mu\nu} d \ra
-4\la \bar{u} d \bar{d} u \ra
+ \la \bar{u}\sigma_{\mu\nu} d \bar{d}\sigma_{\mu\nu} u \ra
\right\}
\right. \nonumber\\[10pt]
& & \left. - {1 \over 2}
\left\{ {\rm with}\ \lambda^a \right\}
\right]\nonumber\\[10pt]
& & + {m_s \over 3q^2} \left[
\left( 1 -{1\over N_c } \right)
\left\{
 \la \bar{u} u \bar{s} s \ra
+ \la \bar{d} d \bar{s} s \ra
-{1 \over 2} \la \bar{u}\gamma_\mu u \bar{s}\gamma_\mu s \ra
-{1 \over 2} \la \bar{d}\gamma_\mu d \bar{s}\gamma_\mu s \ra
\right. \right. \nonumber\\[10pt]
& & \left. \left.
\qquad\qquad\qquad -{1 \over 2} \la \bar{u}\gamma_\mu \gamma_5 u
\bar{s}\gamma_\mu \gamma_5 s \ra
-{1 \over 2} \la \bar{d}\gamma_\mu \gamma_5 d
\bar{s}\gamma_\mu \gamma_5 s \ra
\right\}
\right. \nonumber\\[10pt]
& & \left.
-{1 \over 2}
\left\{ {\rm with}\ \lambda^a \right\}
\right],
\label{a3}\\[15pt]
\tilde{\Pi}^\Lambda_2(q) &=& {-2\over 3q^2} \left[
{1\over 2}\left( 1 -{1\over N_c} \right)
\left\{
{5\over 2} \la \bar{u}\gamma_\mu u \bar{d}\gamma_\mu d \ra
+ {3\over 2} \la \bar{u}\gamma_\mu \gamma_5 u \bar{d}\gamma_\mu \gamma_5 d \ra
+{5\over 4} \la \bar{u}\gamma_\mu u \bar{s}\gamma_\mu s \ra
\right.\right.\nonumber\\[10pt]
& & \left.\left.
\qquad\qquad
+ {3\over 4} \la \bar{u}\gamma_\mu \gamma_5 u \bar{s}\gamma_\mu \gamma_5 s \ra
+{5\over 4} \la \bar{d}\gamma_\mu d \bar{s}\gamma_\mu s \ra
+ {3\over 4} \la \bar{d}\gamma_\mu \gamma_5 d \bar{s}\gamma_\mu \gamma_5 s \ra
\right\}
\right.\nonumber\\[10pt]
& &
\left.
-{1\over 4}
\left\{ {\rm with}\ \lambda^a \right\}
\right.\nonumber\\[10pt]
& & \left.
+\left( 1 - {1\over N_c}\right)
\left\{ \la \bar{u} u \bar{s}s \ra
- \la \bar{u}\gamma_5 u \bar{s}\gamma_5s \ra
+ {1\over 2} \la \bar{u}\gamma_\mu u \bar{s}\gamma_\mu s \ra
- {1\over 2} \la \bar{u}\gamma_\mu\gamma_5 u \bar{s}\gamma_\mu\gamma_5 s \ra
\right.\right.\nonumber\\[10pt]
& & \left.\left.
\qquad\qquad
+ \la \bar{d} d \bar{s}s \ra
- \la \bar{d}\gamma_5 d \bar{s}\gamma_5s \ra
+ {1\over 2} \la \bar{d}\gamma_\mu d \bar{s}\gamma_\mu s \ra
- {1\over 2} \la \bar{d}\gamma_\mu\gamma_5 d \bar{s}\gamma_\mu\gamma_5 s \ra
\right\}
\right.\nonumber\\[10pt]
& &
\left.
-{1\over 2} \left\{ {\rm with}\ \lambda^a \right\}
\right.
\nonumber\\[10pt]
& &
\left.
-{1\over 2}\left( 1 -{1 \over N_c} \right)
\left\{
{5\over 2} \la\bar{u}\gamma_\mu d\bar{d}\gamma_\mu u \ra
+{3\over 2} \la\bar{u}\gamma_\mu\gamma_5 d\bar{d}\gamma_\mu\gamma_5 u \ra
\right\}
+{1\over 4}\left\{ {\rm with}\ \lambda^a \right\} \right].
\label{a4}
\eeq
(iii) $\Sigma$:
\beq
\tilde{\Pi}^\Sigma_1(q) &=& {m_s \over q^2} \left[
\left(1-{1\over N_c}\right)
\left\{
-\la (\bar{u}u)^2 \ra +\la (\bar{u}\gamma_5u)^2 \ra
-{1\over 2}\la (\bar{u}\gamma_\mu u)^2 \ra
+{1\over 2}\la (\bar{u}\gamma_\mu\gamma_5 u)^2 \ra
\right\}
\right.\nonumber\\[10pt]
& &\left. \qquad\qquad
-{1\over 2} \left\{ {\rm with}\ \lambda^a \right\}
\right.
\nonumber\\[10pt]
& & \left. + \left( 1 -{1\over N_c} \right)
\left\{
\la\bar{u}\gamma_\mu u \bar{s}\gamma_\mu s \ra
-\la\bar{u}\gamma_\mu\gamma_5 u \bar{s}\gamma_\mu\gamma_5 s \ra
\right\}
-{1\over 2} \left\{ {\rm with}\ \lambda^a \right\} \right],
\label{a5}\\[15pt]
\tilde{\Pi}^\Sigma_2(q) &=& {-1\over q^2} \left[
\left( 1 -{1\over N_c} \right)
\left\{ {5\over 2} \la\bar{u}\gamma_\mu u \bar{s}\gamma_\mu s \ra
+{3\over 2} \la\bar{u}\gamma_\mu\gamma_5 u \bar{s}\gamma_\mu\gamma_5 s \ra
\right\}
-{1\over 2}\left\{ {\rm with}\ \lambda^a \right\}
\right.\nonumber\\[10pt]
& & \left.+ \left( 1 -{1\over N_c} \right)
\left\{ {1\over 2} \la(\bar{u}u)^2\ra
- {1\over 2} \la(\bar{u}\gamma_5 u)^2\ra
- {1\over 4} \la(\bar{u}\gamma_\mu\gamma_5 u)^2\ra
+ {1\over 4} \la(\bar{u}\gamma_\mu u)^2\ra
\right\}
\right.\nonumber\\[10pt]
& & \left.
\qquad\qquad
-{1\over 2}\left\{ {\rm with}\ \lambda^a \right\} \right].
\label{a6}
\eeq
(iv) $\Xi$:
\beq
\tilde{\Pi}^\Xi_1(q) &=& {m_s \over q^2} \left[
\left(1-{1\over N_c}\right)
\left\{ -3 \la\bar{u}u\bar{s}s\ra
+ \la\bar{u}\sigma_{\mu\nu}u\bar{s}\sigma_{\mu\nu}s\ra
+{1\over 2} \la\bar{u}\sigma_{\mu\nu}\gamma_5 u
\bar{s}\sigma_{\mu\nu}\gamma_5 s\ra
\right\}
\right.\nonumber\\[10pt]
& & \left. \qquad\qquad
-{1\over 2} \left\{ {\rm with}\ \lambda^a \right\} \right],
\label{a7}\\[15pt]
\tilde{\Pi}_2^\Xi(q) &=& {-1\over q^2} \left[
\left( 1 -{1\over N_c} \right)
\left\{
{5\over 2} \la\bar{u}\gamma_\mu u\bar{s}\gamma_\mu s \ra
+{3\over 2} \la\bar{u}\gamma_\mu\gamma_5 u\bar{s}\gamma_\mu\gamma_5 s \ra
\right\} - {1\over 2}\left\{ {\rm with}\ \lambda^a \right\}
\right.\nonumber\\[10pt]
& & \left.
+\left( 1 - {1\over N_c} \right)
\left\{
{1\over 2} \la (\bar{s}s)^2\ra
-{1\over 2} \la (\bar{s}\gamma_5 s)^2\ra
-{1\over 4} \la (\bar{s}\gamma_\mu \gamma_5 s)^2\ra
+{1\over 4} \la (\bar{s}\gamma_\mu s)^2\ra
\right\}
\right.\nonumber\\[10pt]
& & \left. \qquad\qquad
-{1\over 2} \left\{ {\rm with}\ \lambda^a \right\} \right].
\label{a8}
\eeq
Factorizing these four-quark operators into the square of the
chiral order parameter, one can easily get the form of
(\ref{eq209}), (\ref{eq210}) and the following (\ref{eq211})--(\ref{eq215})
at $T=0$.  To get the $T$-dependence of the four-quark condensates,
we need to calculate the double commutators of (\ref{a1})--(\ref{a8})
with the isovector axial charge.
The calculation is tedious but straightforward.
The following formulas are useful in carrying out the calculation ($q=(u,d)$):
\beq
\left[{\cal F}_5^a,\bar{q}q\right]&=&-\bar{q}\gamma_5 \tau^a q,\nonumber\\[8pt]
\left[{\cal F}_5^a,\bar{q}\gamma_5q\right]&=&-\bar{q}\tau^a q,\nonumber\\[8pt]
\left[{\cal F}_5^a,\bar{q}\sigma_{\mu\nu}q\right]
&=&-\bar{q}\sigma_{\mu\nu}\gamma_5\tau^a q,\nonumber\\[8pt]
\left[{\cal F}_5^a,\bar{q}\sigma_{\mu\nu}\gamma_5q\right]
&=&-\bar{q}\sigma_{\mu\nu}\tau^a q,\nonumber\\[8pt]
\left[{\cal F}_5^a,\bar{q}\gamma_\mu q\right]&=&
\left[{\cal F}_5^a,\bar{q}\gamma_\mu\gamma_5 q\right]=0,\nonumber\\[8pt]
\left[{\cal F}_5^a,\bar{q}\tau^b q\right]
&=&-\delta^{ab}\bar{q}\gamma_5 q,\nonumber\\[8pt]
\left[{\cal F}_5^a,\bar{q}\gamma_5\tau^b q\right]
&=&-\delta^{ab}\bar{q}q,\nonumber\\[8pt]
\left[{\cal F}_5^a,\bar{q}\sigma_{\mu\nu}\tau^b q\right]
&=&-\delta^{ab}\bar{q}\sigma_{\mu\nu}\gamma_5 q,\nonumber\\[8pt]
\left[{\cal F}_5^a,\bar{q}\sigma_{\mu\nu}\gamma_5\tau^b q\right]
&=&-\delta^{ab}\bar{q}\sigma_{\mu\nu}q,\nonumber\\[8pt]
\left[{\cal F}_5^a,\bar{q}\gamma_\mu\tau^b q\right]
&=&i\epsilon^{abc}\bar{q}\gamma_\mu\gamma_5 \tau^c q,\nonumber\\[8pt]
\left[{\cal F}_5^a,\bar{q}\gamma_\mu\gamma_5\tau^b q\right]
&=&i\epsilon^{abc}\bar{q}\gamma_\mu \tau^c q.
\eeq
After calculating the double commutators, we end up with other four-quark
operators.  Applying the factorization to these four-quark operators,
we eventually got (\ref{eq209}), (\ref{eq210}) and the following
(\ref{eq211})--(\ref{eq215}):
\beq
\Pi^\Lambda_1(q,T) &=& {1 \over 12\pi^2}\left(4\la\bar{u}u\ra
\left(1-{\zeta\over 8}\right)
-\la\bar{s}s\ra\right)q^2{\rm ln}(Q^2) + {m_s \over 96\pi^4}q^4{\rm ln}(Q^2)
\nonumber\\[10pt]
& &-{4m_s \over 3q^2}\la\bar{u}u\ra^2\left(1-{\zeta \over 6}\right)
+{4m_s \over 9q^2}\la\bar{u}u\ra\la\bar{s}s\ra \left(1-{\zeta \over 8}\right),
\label{eq211}
\\[10pt]
\Pi^\Lambda_2(q,T) &=& {-1\over 64\pi^4}q^4{\rm ln}(Q^2)
+{m_s\over 12\pi^2}\left(4\la\bar{u}u\ra\left(1-{\zeta\over 8}\right)
-3\la\bar{s}s\ra\right){\rm ln}(Q^2)\nonumber\\[10pt]
& &-{1 \over 32\pi^2}
\la{\alpha_s \over \pi} G^2\ra{\rm ln}(Q^2)
+{2\over 9q^2}\la\bar{u}u\ra^2\left(1-{\zeta\over 2}\right)
-{8\over 9q^2}\la\bar{u}u\ra\la\bar{s}s\ra\left(1-{\zeta\over 8}\right),
\label{eq211a}\\[15pt]
\Pi^\Sigma_1(q,T)&=&{1\over 4\pi^2}\la\bar{s}s\ra q^2 {\rm ln}(Q^2)
-{m_s \over 32\pi^2}q^4{\rm ln}(Q^2) -{4m_s \over 3q^2}\la\bar{u}u\ra^2
\left(1-{\zeta\over 6}\right),\label{eq212}
\\[10pt]
\Pi^\Sigma_2(q,T)&=&{-1\over 64\pi^4}q^4{\rm ln}(Q^2)
-{m_s \over 4\pi^2}\la\bar{s}s\ra{\rm ln}(Q^2)\nonumber\\[10pt]
& &-{1 \over 32\pi^2}
\la{\alpha_s \over \pi} G^2\ra{\rm ln}(Q^2)
-{2\la\bar{u}u\ra^2\over 3q^2}\left(1-{\zeta\over 8}\right),\label{eq213}
\\[15pt]
\Pi^\Xi_1(q,T)&=&{1\over 4\pi^2}\la\bar{u}u\ra\left(1-{\zeta\over 8}\right)
-{2m_s\over q^2}\la\bar{u}u\ra\la\bar{s}s\ra\left(1-{\zeta\over 8}\right),
\label{eq214}
\\[10pt]
\Pi^\Xi_2(q,T)&=&{-1\over 64\pi^4}q^4{\rm ln}(Q^2)
-{1 \over 32\pi^2}
\la{\alpha_s \over \pi} G^2\ra{\rm ln}(Q^2)
-{2\la\bar{s}s\ra^2\over 3q^2}.
\label{eq215}
\eeq
The $T$-dependence of (\ref{eq214}) and (\ref{eq215}) is the same as
(\ref{eq410}).  We also note the $T$-dependence of the four-quark operators
is different in the different channels.

\newpage

\centerline{\bf Figure Captions}
\vskip 15pt
\begin{description}

\item[Fig. 1] $\pi+N \rightarrow N$ scattering term in which
$\eta^N$ couples to $\pi$ directly.

\item[Fig. 2] $\pi+N \rightarrow N$ scattering term in which
$\eta^N$ couples to the nucleon that interacts with $\pi$.

\end{description}
\end{document}